\documentclass[12pt]{article}

\usepackage{graphicx}
\newcommand{\sig}{\mbox{\boldmath $\sigma$}}
\newlength{\gmm} 
\begin{document}

\title{The Dirac Equation in Classical Statistical Mechanics}
\author{G. N. Ord\\
  M.P.C.S. \\ Ryerson University\\Toronto Ont.}

\maketitle

\begin{abstract}
The Dirac equation, usually obtained by `quantizing' a classical
stochastic model is here obtained directly within classical statistical
mechanics. The special underlying space-time geometry of the random walk
replaces the missing analytic continuation, making the model
`self-quantizing'. This provides a new context for the Dirac equation,
distinct from its usual context in relativistic quantum mechanics.
\end{abstract}

\maketitle

%%%%%%%%%%%%%%%%%%%%%%%%%%%%%%%%%%%%%%%%%%%%
%% MAINMATTER
%%%%%%%%%%%%%%%%%%%%%%%%%%%%%%%%%%%%%%%%%%%%

\section{Introduction}
 The title of this talk requires some explanation. Statistical
mechanics is in some respects the simplest branch of physics; all you ever
have to do  is  count.  The trick is of course to 
find the right class of objects to count. 

The implication of the title is that the Dirac equation is simply related to the operation of counting objects.
If we were talking about the diffusion or heat equation then there would be no mystery. It is
well known  that the diffusion equation may be obtained by counting random walks in an appropriate limit.

However, the likelihood of being able to make a similar claim for the Dirac equation seems small. We all know that
the Dirac equation has elements of quantum mechanics, special relativity and
half-integral spin. None-the-less the claim in the title is true for the
Dirac equation in one dimension
\cite{gord01a}, and may well be true in  three dimensions
 for a free particle. Among other things, this means that the Dirac
equation has a context as a phenomenology which may be distinct from its
role as a fundamental equation of quantum mechanics. Although this may be surprising from the perspective of
quantum mechanics, it is not so extraordinary in the context of general partial differential equations. For example
we are used to calling the same PDE either the Diffusion equation or the heat equation, depending on context. In
the case of Dirac, we have only one name for the equation, but there are still multiple contexts.

 Today I am going to talk about the Dirac equation  as a classical phenomenology. However since we all think
`quantum mechanics' at the mention of Dirac's name, let's recall  where Quantum mechanics begins and where it
ends. Typically we pass from classical physics to
quantum mechanics through a formal analytic continuation (FAC).
The canonical FAC is to replace the momentum by the operator $-i\hbar
\nabla$ and the energy by $i \hbar \frac{\partial}{\partial t}$. In a
sense this is where quantum mechanics begins and classical physics ends. 

The quantum equations describe the evolution of the initial conditions in
time, and we relate the results of this evolution to our macroscopic world
by the measurement postulates. These postulates are the weakest link in
the theory, and it is here that the (many) interpretations of quantum
mechanics vie for supremacy. At this point we do not really know where
quantum mechanics ends and classical physics begins. So let us return to
the point where it begins, at the FAC.

The the canonical FAC just mentioned specifically relates the Schr\"{o}dinger
equation to Hamiltonian mechanics. We will not be so specific.
Table(1)  compares classical PDE's with their corresponding
`quantum' counterparts. In the left `Classical' column we start out with a
two component form of the Telegraph equations due to Marc Kac. Here U is related to a
two component probability density, $c$ is a mean free speed and $a$ is an
inverse mean free path. Note that if we replace the real positive constant
$a$ by $im$ we get a form of the Dirac equation.

\begin{table}
\begin{tabular}{crr}
\hline
  & {\bf   Classical }
  & {\bf  Quantum }  
     \\

\hline
 {\bf Microscopic  basis}	    &  Kac (Poisson)  & Chessboard     \\
 First Order & \(\frac{\partial {\bf U}}{\partial t}= c \sig_z
                         \frac{\partial
                      \mathbf U}{\partial z} + a \sig_x \mathbf U\) & $\frac{\partial
                      \mathbf
                      \Psi}{\partial t}= c\, \sig_z  \frac{ \partial }{\partial z} \mathbf
                      \Psi +i  \, m\sig_x \mathbf \Psi$    \\
 Second Order & $ \frac{\partial^2 U}{\partial t^2} =
c^2\frac{\partial^2 U}{\partial z^2}
                       +a^2 U$ & $ \frac{\partial^2 \psi}{\partial t^2} =
                       c^2\frac{\partial^2
                      \psi}{\partial z^2}  -  m^2 \psi$   \\
 `Non-relativistic' & $\frac{\partial U}{\partial t}=D \, \frac{\partial^2 U}{\partial
x^2}$ & $\frac{\partial \psi}{\partial t}=  i \,D \,\frac{\partial^2
\psi}{\partial x^2}$  \\
\hline
\end{tabular}
\caption{The relation between Classical PDE's based on stochastic models, and their `Quantum' cousins. FAC
 accomplishes the trick, but then the stochastic basis for the equations becomes formal. }
\label{tab:a}
\end{table}
Similarly the second order form of the Telegraph equations continues to the
Klein-Gordon or Relativistic-Schr\"{o}dinger equation using the same FAC.
The `non-relativistic limit' of the Telegraph equation gives the Diffusion
equation, with the usual FAC to  the Schr\"{o}dinger equation.
The equations on the left are phenomenologies which are interpreted through
the underlying statistical mechanical models (Poisson or Brownian motion).
The equations on the right are regarded as fundamental equations which have
no realistic microscopic basis, and are interpreted through postulates. The analog of Poisson paths for the
Telegraph equations are the Chessboard paths of Feynman.

What we will do today is to show that the Dirac equation in 1-dimension is
also a phenomenological equation with a microscopic basis, and is
accessible directly through classical statistical mechanics \ldots all we
have to do is to find the right objects to count.  But first let us see how
Kac\cite{Kac74}
 obtained the Telegraph equations from a microscopic model.

\section{The Kac Model}

Imagine a particle on a discrete space-time lattice with spacings $\Delta
z$ and $\Delta t$. The speed of particles are fixed at $c$ and occasionally they
scatter backwards with probability $\alpha \Delta t$. Fig. (1.A) shows a typical Kac path. Considering the
density of particles moving in the plus and minus directions  we can write
difference equations for their conservation.
\begin{equation}
F_n^+(z) = (1-a\Delta t) F_{n-1}^+(z-c\Delta t) \label{diff1}
  + a \Delta t
 F_{n-1}^-(z)
\end{equation}
Similarly,
\begin{equation}
F_n^-(z) = (1-a\Delta t) F_{n-1}^-(z+c\Delta t)\label{diff2}
  + a \Delta t
 F_{n-1}^+(z)
\end{equation}

In the continuum limit these give the coupled PDE's:
\begin{equation}
\frac{\partial F^+}{\partial t}= -c \frac{\partial F^+}{\partial z} - aF^+
+aF^- \label{tg1}
\end{equation}

\begin{equation}
\frac{\partial F^-}{\partial t}=c \frac{\partial F^-}{\partial z} + aF^+
- aF^- \label{tg2}
\end{equation}
It is easy to identify the streaming and scattering terms here, but if we
remove the exponential decay, and write in 2-component form we get.

\begin{equation}
\frac{\partial \mathbf U}{\partial t}= c \sig_z \frac{\partial \mathbf
U}{\partial z} + a \sig_x \mathbf U \label{tgnoexp}
\end{equation}
where the \sig  \, are the usual Pauli Matrices.

Note the suggestive form of the coupled equations. We are a FAC away from the Dirac equation.
However the FAC destroys any  coherent interpretation of our microscopic model, so we will
strictly avoid  it!
\begin{figure}
\includegraphics[scale = .42]{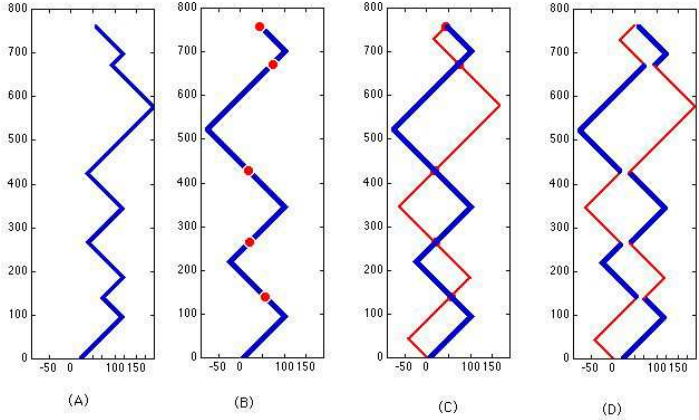}% needs 
\caption{(A) Paths in the Kac process are broken-line segments with slope $\pm c$. The average distance between
corners is $1/a$.(B) The Kac Stochastic process with a stutter. (C) With the return Path (indicated by the
thinner line). (D) The outer envelopes of the Entwined paths.  } 
\end{figure}

\section{The Entwined Pair Model}

As mentioned above, counting Kac Paths gives the telegraph equations.
So what do we count if we want to go directly to the Dirac equation?
And how is $i$ going to appear if we are not allowed a FAC? To answer the second question first, 
we do not need $i$ to get the quantum equations,
we only need the fact that $i^2$=-1. This may seem like a trivial point but
it is extremely important. To illustrate it, think of the even and odd
parts of
$e^t$
\ldots these are $\cosh(t)$ and $\sinh(t)$. Now analytically continue and
think of the even and odd parts of $e^{it}$. These are $\cos(t)$ and $i
\sin(t)$. It is true that here $i$ distinguishes the even and odd
parts, but the important effect of the analytic continuation is to produce
the alternating signs in the series expansion of the trigonometric functions. It is the same
in the quantum equations. We do not actually need $i$, what we need is the alternating pattern of sign that 
gives oscillatory behaviour, and we have to get it from the geometry of particle trajectories, not from a FAC.

Consider the Entwined paths of Fig(1. B-D). We will generate these with the same process we used to generate the
Kac paths of Fig.(1. A), except here we will employ a periodic `stutter'. That is, wherever we would change
direction in a Kac path, we alternately change direction or leave a marker in the entwined path. At the first
marker past some specified time $t_R$, we reverse our direction in $t$ and follow the markers back to the
origin. In Fig(1. C),  the return path has a thinner line-width to distinguish it from the forward path. Even in
Classical physics, particles which move backwards in time behave like anti-particles to a forward moving
observer, so if we record  (+1) as a charge carried on  forward portions of the trajectory, we shall associate a -1
with the return portions. The task will then be to calculate the expected average charge deposited by an ensemble
of these paths. Note that  each entwined pair can be regarded as two osculating envelopes with a periodic
colouring(Fig.1. D). Each envelope is just a Kac path. Each has the same statistics and geometry as a Kac path,
the only difference is the periodic colouring. We can then  set up a difference equation as  we did for the
Telegraph equations. This time there are 4 states instead of Kac's 2 states. Fig.(2. A) shows the four possible
states of an entwined pair and Fig.(2. B) shows a path evolving through two loops.

\begin{figure}
\setlength{\unitlength}{.5mm}
\newlength{\gnat} 
\setlength{\gnat}{45mm}
\begin{picture}(70,80)(-25,-5)
\put(45 ,4){$( -1,0, 0,1)$}% State label of first state
\thicklines 
\put(10,0){\thinlines{\vector(-1,1){10}}}
\put(20,0){ \thicklines{\vector(1,1){10}}}
\put(45,24){$( 0,-1,0,1)$} % State label of second state
\put(0,20){ \thinlines{\vector(1,1){10}}}
\put(20,20){ \thicklines{\vector(1,1){10}}}
\put(45,44){$(-1,0, 1,0)$} % State label of third state
\put(10,40){\thinlines{\vector(-1,1){10}}}
\put(30,40){ \thicklines{\vector(-1,1){10}}}
\put(45,64){$( 0,-1, 1,0)$} % State label of first state 
\put(0,60){ \thinlines{\vector(1,1){10}}}
\put(30,60){ \thicklines{\vector(-1,1){10}}}
\put(-5,-5){\framebox(100,80)}
\end{picture}\hspace{3cm}
\begin{picture}(70,80)(-25,-5)
\put(45 ,4){$( -1,0, 0,1)$}% State label of first state
\thicklines 
\put(10,0){{\thinlines\vector(-1,1){10}}}
\put(10,0){ \thicklines{\vector(1,1){10}}}
\put(45,14){$( 0,-1,0,1)$} % State label of second state
\put(0,10){ \thinlines{\vector(1,1){10}}}
\put(20,10){ \thicklines{\vector(1,1){10}}}
\put(45,24){$(0,-1, 1,0)$} % State label of third state
\put(10,20){ \thinlines{\vector(1,1){10}}}
\put(30,20){ \thicklines{\vector(-1,1){10}}}
\put(45,34){$( 1,0, 0,-1)$} % State label of first state 
\put(20,30){ \thinlines{\vector(1,1){10}}}
\put(20,30){ \thicklines{\vector(-1,1){10}}}
\put(45,44){$( 1,0, -1,0)$} % State label of first state 
\put(30,40){ \thinlines{\vector(-1,1){10}}}
\put(10,40){ \thicklines{\vector(-1,1){10}}}
\put(45,54){$( 0,1, -1,0)$} % State label of first state 
\put(20,50){ \thinlines{\vector(-1,1){10}}}
\put(0,50){ \thicklines{\vector(1,1){10}}}
\put(-5,-5){\framebox(110,75)}
\end{picture}
\caption{ The four possible states of Entwined Pairs and two cycles of an Entwined Pair. Note the alternating
signs in the envelope.}
\end{figure}
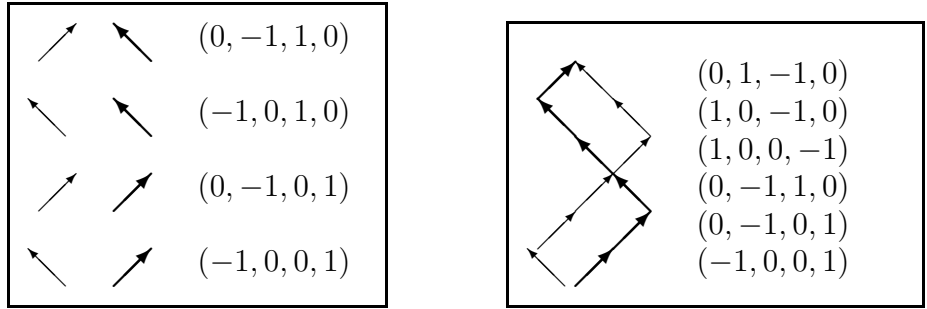
%Here are the difference equations for the entwined paths.
\begin{eqnarray} \nonumber
\phi_n^1(z)& =& (1-a\Delta t) \phi_{n-1}^1(z-c\Delta t)
  - a \Delta t
 \phi_{n-1}^2(z-c\Delta t)\\ \nonumber
\phi_n^2(z)& =& (1-a\Delta t) \phi_{n-1}^2(z+c\Delta t) 
  + a \Delta t
 \phi_{n-1}^1(z+c\Delta t)\\
\phi_n^3(z)& =& (1-a\Delta t) \phi_{n-1}^3(z-c\Delta t) \label{dirdiff1}
  - a \Delta t
 \phi_{n-1}^4(z-c\Delta t)\\\nonumber
\phi_n^4(z)& =& (1-a\Delta t) \phi_{n-1}^4(z+c\Delta t) 
  + a \Delta t
 \phi_{n-1}^3(z+c\Delta t)
\end{eqnarray}
 Note the alternating minus sign from the crossover of paths (cf. Feynman Chessboard Model).
Here the $\phi$ are real ensemble averages of a net `charge' density. They are {\em  not} quantum mechanical
amplitudes. 

Removing the exponential decay and writing $p_z=-i\frac{\partial}{\partial z}$, setting $c = 1$ and
$a =m$ with
\(
\alpha_z=\left (\begin{array}{c c}
-\sigma_z &0\\
0&\sigma_z
\end{array}\right )
\)  \(
\beta=\left (\begin{array}{c c}
\sigma_y &0\\
0&\sigma_y
\end{array}\right )
\)
we get the Dirac equation

\begin{equation}
i\frac{\partial \psi}{\partial t}=(\alpha_z . p_z +\beta m)\psi
\label{dirac1}
\end{equation}
in the continuum limit. 
Notice there is no FAC here. The $i$ in Eqn.(\ref{dirac1}) was introduced simply to show a familiar form of the
Dirac equation. The 
$\psi$ here are real, four component, and oscillatory in character. The oscillation is implemented through the
presence of
\(\sigma_q= i\sigma_y
\), which is a real, anti-hermitian matrix. $\sigma_q$ arises because of the periodic
exchange of particle and antiparticle in entwined paths, and it has the important feature
that $\sigma_q^2=-1$. We have not forced an analytic continuation on the system here. The
space-time geometry itself has rendered the system `self-quantizing'!

\section{Discussion}

The above suggests that the Dirac equation appears as an ensemble average of a net charge over our entwined pairs.
Since all pairs meet at the origin the entire ensemble can be regarded as being generated {\em by a single
particle traversing all entwined paths}. However the above analysis only shows that the Dirac propagator will
result if we cover the ensemble exactly. There have been other cases where the quantum equations have been
recovered as projections without using a FAC
\cite{gord92,gord93lett,gord96,McKeonOrd92,gordDeakin97,gordDeakin96}, however all of these have required a
complete ensemble. What happens if we just watch the stochastic process with its inherent  fluctuations? Will the
process converge to the propagator or will the fluctuations swamp the signal? The answer appears to be that the
signal survives stochastic fluctuations. The Dirac propagator is a stable feature of entwined paths!  Figures
(3) and (4) show the propagator drawn by a single path in  comparison to the discrete Dirac
propagator.\cite{gord01a}
\begin{figure}
\includegraphics[scale = .55]{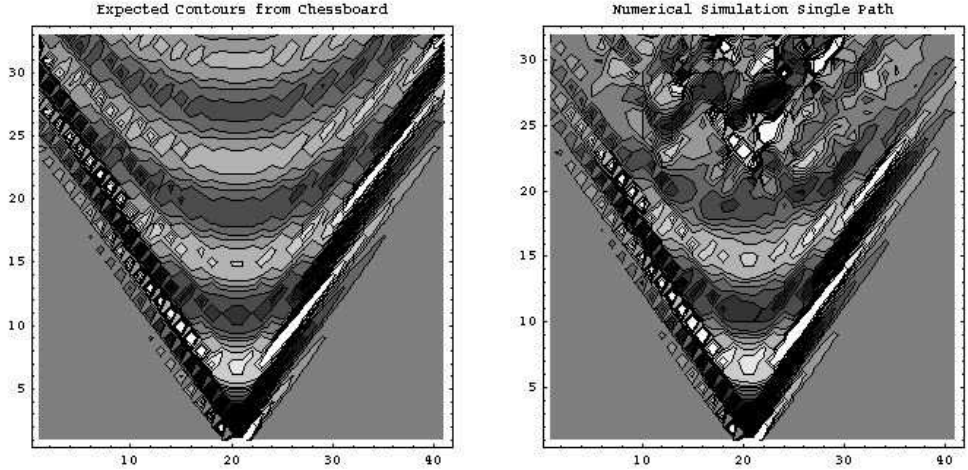}% needs 
\caption{Contour plot of the sum of the real and imaginary parts of the Dirac Propagator. Left, discrete Dirac,
right, single path.} 
\end{figure}
\begin{figure}
\includegraphics[scale =.99]{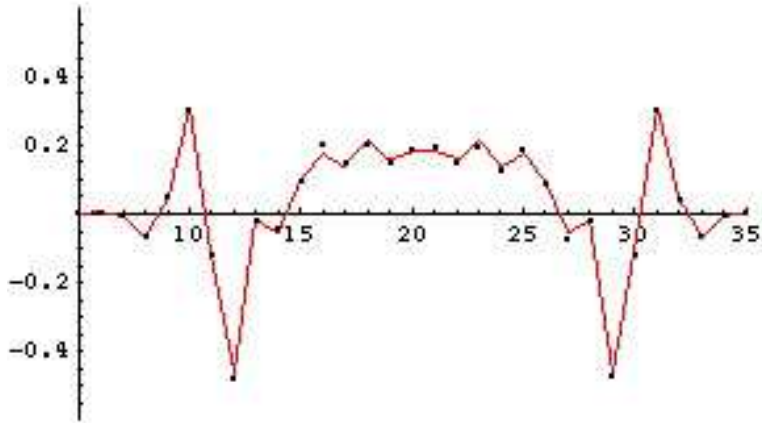}% needs 
\caption{A time slice through the above figure at $t=15$. The  curve is the Dirac propagator and the black
dots are the single path.} 
\end{figure}
In Fig.(3) we see that the propagator formed by the entwined paths is quite accurate near the origin but gets
less accurate the father out we go. This is to be expected since the configuration space grows
exponentially as the number of time steps and the farther away we are from the origin, the poorer the ensemble
coverage.

Considering the original context of the Dirac Equation, the above demonstration that the equation is easily
obtained within classical statistical mechanics is surprising, to say the least. Clearly, the algebraic
route followed by Dirac to obtain his equation was, and is, elegant, concise and algebraically compelling.
However, for all its comparative inelegance, the heuristic approach via entwined paths strongly suggests that
canonical quantization, Dirac's starting point, might well benefit from a re-evaluation. The FAC represented by
canonical quantization gives an algebraic connection to Hamiltonian mechanics without any suggestion as to what
the connection actually means in terms of the propagation of a classical particle. By comparison, the entwined
paths approach  replaces a formal algebraic requirement by {\em a physical constraint on space-time geometry}.
What we lose in elegance we may well regain in  physical cogency. As Dirac suggested in the  preface to his
book\cite{dirac58}

\begin{quote}

  ``Mathematics is the tool specially suited for dealing with abstract
concepts,
\ldots All the same, the mathematics is only a tool and one should learn
to hold the physical ideas in one's mind without reference to the
mathematical form.'' 

\end{quote}
%%%%%%%%%%%%%%%%%%%%%%%%%%%%%%%%%%%%%%%%%%%%%%%%

\end{document}